\documentclass[aps,prl,reprint,groupedaddress]{revtex4-1}

\usepackage{graphicx}
\usepackage{dcolumn}
\usepackage{bm}

\newcommand{\fig}[4][t]{\begin{figure}[#1]\begin{center}\includegraphics[scale=#2]{#3.pdf}\vspace{-0.25 cm}\caption{#4}\label{fig:#3}\end{center}\end{figure}}

\newcommand{\etal}[0]{\emph{et al.}}

\begin{document}

\title{Antiresonance phase shift in strongly coupled cavity QED}

\author{C. Sames, H. Chibani, C. Hamsen, P.~A. Altin, T. Wilk and G. Rempe}
\email{gerhard.rempe@mpq.mpg.de}
\affiliation{Max-Planck-Institut f\"ur Quantenoptik, Hans-Kopfermann-Str.\ 1, D-85748 Garching, Germany}

\date{\today}

\begin{abstract}
We investigate phase shifts in the strong coupling regime of single-atom cavity quantum electrodynamics (QED). On the light transmitted through the system, we observe a phase shift associated with an antiresonance and show that both its frequency and width depend solely on the atom, despite the strong coupling to the cavity. This shift is optically controllable and reaches 140$^\circ$ -- the largest ever reported for a single emitter. Our result offers a new technique for the characterization of complex integrated quantum circuits.
\end{abstract}


\maketitle


The strongly coupled atom-cavity system plays a central role in research on fundamental quantum optics. Important achievements to date include the creation of single photon sources \cite{Kuhn:2002,McKeever:2004} and non-classical microwave states \cite{Deleglise:2008,Guerlin:2007}, single-atom squeezing \cite{Ourjoumtsev:2011}, the observation of novel photon statistics \cite{Birnbaum:2005,Kubanek:2008,Koch:2011} and the nondestructive detection of microwave and optical photons \cite{Nogues:1999,Reiserer:2013}. More complex interacting systems based on this basic element are now attracting much attention in quantum information and simulation. Recent achievements in this direction include the coupling of a single qubit to two cavities \cite{Kirchmair:2013}, the interaction of multiple qubits with a single cavity bus \cite{Majer:2007,Mariantoni:2011}, and the exchange of quantum states between single qubits in remote cavities \cite{Ritter:2012,Nolleke:2013}. Integrated quantum circuits are promising candidates for on-chip quantum computation \cite{Ladd:2010,Monroe:2013,Awschalom:2013,Politi:2008,Benson:2011} and large strongly coupled networks have been proposed for simulating quantum phase transitions \cite{Greentree:2006,Hartmann:2006,Houck:2012}. However, in such strongly interacting systems, the couplings no longer represent merely a perturbation of the subsystem dynamics, necessitating a holistic analysis of the coupled system. This makes the characterization of strongly coupled quantum circuits a challenging task \cite{Devoret:2013,Nigg:2012}.

In this Letter, we propose a new technique for characterizing complex quantum circuits, which emerges from an analysis of the phase of light transmitted through a strongly coupled single-atom--cavity system. In particular, we report on the observation of an antiresonant phase shift caused by destructive interference between the coherent drive and the field radiated by the atom. The signature of the antiresonance is a large negative phase shift which depends solely on the atom, despite the strong coupling to the resonator. This is in sharp contrast to the normal modes \cite{Boca:2004,Maunz:2005}, which depend on properties of both atom and cavity as well as the coupling strength \cite{Raizen:1989}. Our measurement paves the way for individual components of strongly interacting quantum systems to be characterized via measurements performed only on the overall coupled system.


Previous work on phase spectroscopy in cavity QED has focused on the so-called ``bad-cavity'' limit in which the cavity decay rate exceeds the coupling strength, $\kappa \gtrsim g$, and only modest phase shifts were observed \cite{Turchette:1995,Fushman:2008}. Phase changes due to strongly coupled atoms were seen in Ref.\ \cite{Mabuchi:1999}, but the antiresonance phase shift was not observed. The presence of a transmission dip at the atomic frequency (associated with the antiresonance) was noted in theoretical work in the intermediate-coupling limit \cite{Rice:1996} \footnote{The work of Ref. {\cite{Rice:1996}} likened the transmission dip in the intermediate-coupling regime to electromagnetically induced transparency. However, we note that the phase shift studied in this work cannot be observed on the light transmitted through an EIT medium, despite the apparent similarity between Eq.\ (1) and the EIT susceptibility $\chi$. This is because the phase acquired by light passing through an EIT medium is proportional to $\Re(\chi)$ (the refractive index), not to its argument. Here, we measure directly the phase of {$\langle \hat{a} \rangle$}.}. In contrast, in a strongly interacting system the coupling exceeds all decay rates, such that excitations are coherently exchanged between atom and cavity, leading to the formation of a new set of eigenstates. In this limit, the antiresonance occurs far from these new eigenstates, which impedes its observation via the intensity transmitted through the cavity \cite{Boca:2004,Maunz:2005}. Here we clearly reveal the antiresonant behavior through a measurement of phase.


In the limit of low atomic excitation, the expectation value of the cavity field (represented by the photon annihilation operator $\hat{a}$) can be straightforwardly calculated within the framework of the Jaynes-Cummings model, extended to take into account driving and dissipation:
\begin{equation}
\langle\hat{a}\rangle = \frac{ \eta ( \Delta_{pa} + i\gamma ) } {( \Delta_{pa} + i\gamma ) ( \Delta_{pc} + i\kappa ) - g^2} \,,
\end{equation}
where $\gamma$ denotes the atomic dipole decay rate, $\eta$ is the strength of the coherent drive, and $\Delta_{pa} = \omega_p - \omega_a$ and $\Delta_{pc} = \omega_p - \omega_c$ represent the probe-atom and probe-cavity frequency detunings, respectively. The antiresonance phase shift in this system occurs when the numerator of Eq.\ (1) is minimized, at $\Delta_{pa} = 0$. Remarkably, this depends only on atomic parameters; the antiresonance occurs at exactly the resonance frequency $\omega_a$ of the uncoupled atom, and has a width equal to the bare atomic linewidth $\gamma$, despite the strong coupling between atom and cavity. If the roles of atom and cavity are exchanged by driving the atom at the empty-cavity resonance, the steady-state light field in the cavity reaches a magnitude equal to the drive, such that the atom then remains in its ground state \cite{Alsing:1992,Zippilli:2004}.


Our strongly coupled system consists of a single $^{85}$Rb atom ($\gamma/2\pi = 3.0$\,MHz) in a high-finesse ($\mathcal{F} = 195,000$) Fabry-Perot cavity of length 260\,$\mu$m ($\kappa/2\pi = 1.5$\,MHz). An atom-cavity coupling constant of $g_0/2\pi = 16$\,MHz at an antinode of the cavity field puts the system well into the strong coupling regime of cavity QED, $g \gg (\gamma,\kappa)$. A circularly polarized laser beam at 785\,nm serves as an intra-cavity dipole trap for single atoms and is used to actively stabilize the cavity length, which matches a frequency $\omega_c$ blue detuned by 40\,MHz from the $5S_{1/2}, F=3\rightarrow5P_{3/2},F=4$ cycling transition. The ac-Stark shift caused by the dipole trap then results in an atom-cavity detuning $\Delta_{ac}=\omega_a - \omega_c$ of only a few MHz. Heterodyne detection is used to probe the magnitude and phase of the light field transmitted through the coupled system (see Figure 1).

\fig{0.32}{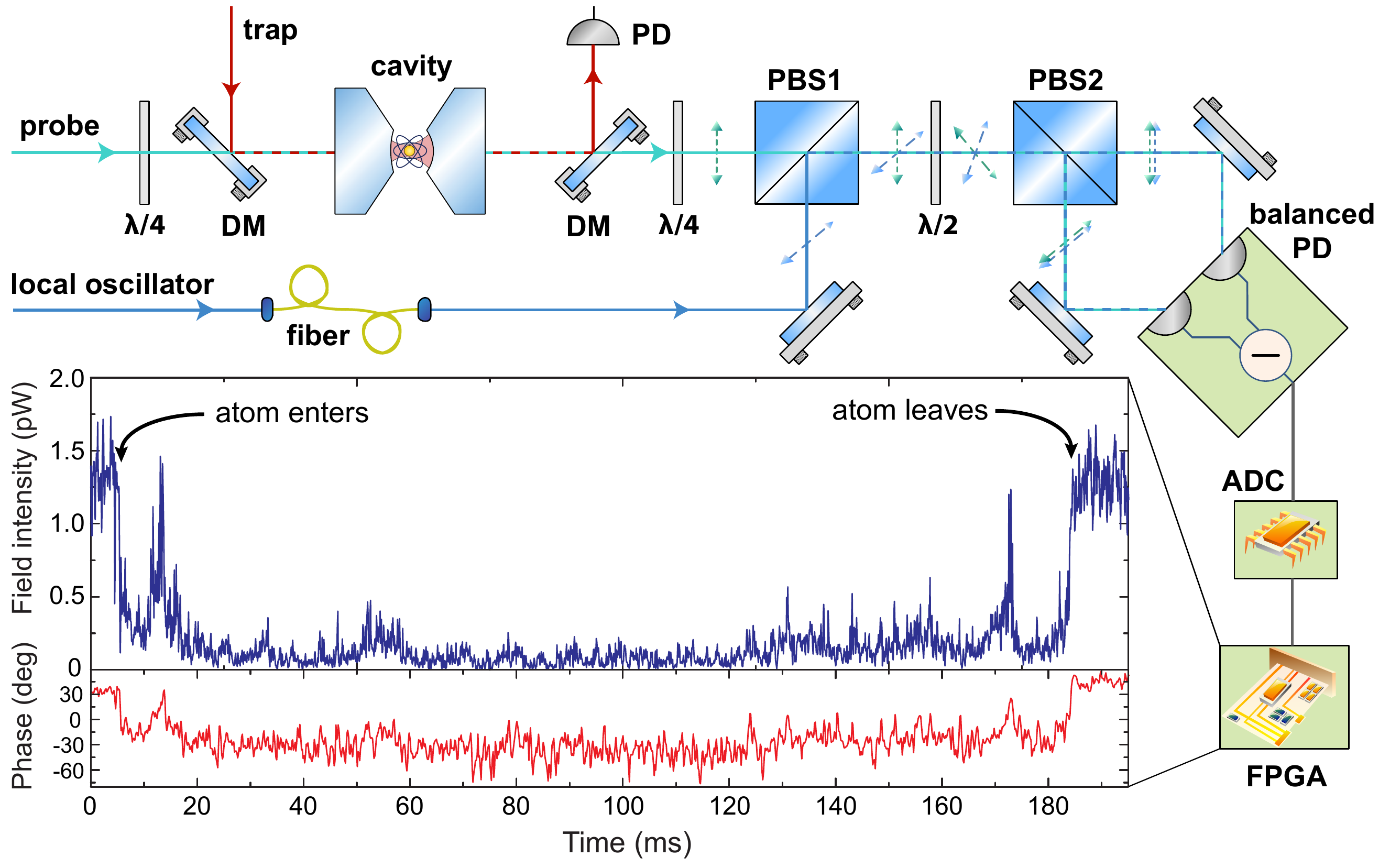}{Experimental setup. The circularly polarized probe beam transmitted through the cavity is changed to linear before being overlapped on polarizing beam splitter PBS1 with a local oscillator of orthogonal polarization. The subsequent $\lambda/2$-waveplate rotates the polarization of both beams by $45^\circ$ so that they are split equally at PBS2 before photodetection (PD). The difference of the photocurrents from the two arms is digitized by an analog-to-digital converter (ADC) and fed into a field-programmable gate array (FPGA), which reconstructs the amplitude and phase information. The dipole-trap and probe beams are merged and separated by dichroic mirrors (DM). A typical trace is shown in the inset for an atom held in the intracavity dipole trap for 180\,ms. The drop in the field intensity heralds the presence of an atom in the cavity.}

The atom is loaded into the cavity by means of a pulsed atomic fountain. A drop in the amplitude of a resonant probe beam ($\Delta_{pc} = 0$) heralds the arrival of an atom, triggering an increase in the power of the dipole laser and thus capturing the atom. In order to prepare the system reliably in the strong coupling regime, probe intervals are interleaved with cooling intervals, in which cavity and feedback cooling are applied \cite{Maunz:2004,Kubanek:2009,Koch:2010}. During probe intervals, the feedback algorithm is disabled and the trap is kept at a constant value. The frequency of the probe beam is then changed to the value under study and its intensity decreased to avoid heating and saturation of the atom \footnote{See Supplemental Material for details of cooling and confinement, heterodyne detection, phase shift measurement and numerical simulation.}.


Theoretical amplitude and phase spectra for our atom-cavity parameters are shown in Figure 2a,b. The black lines represent the frequency response of the empty cavity. The response changes significantly when an atom is strongly coupled to the cavity mode, resulting in the appearance of normal modes (denoted $| 1,- \rangle$ and $| 1,+ \rangle$) where the excitation is shared between the atom (green) and the cavity (red). In a logarithmic plot of the cavity excitation, a dip (the antiresonance) becomes evident at the resonance frequency of the bare atom. No feature is apparent in either the atomic excitation or phase at this frequency, which demonstrates that the effect is not merely interference between the two normal modes.

\fig{0.4}{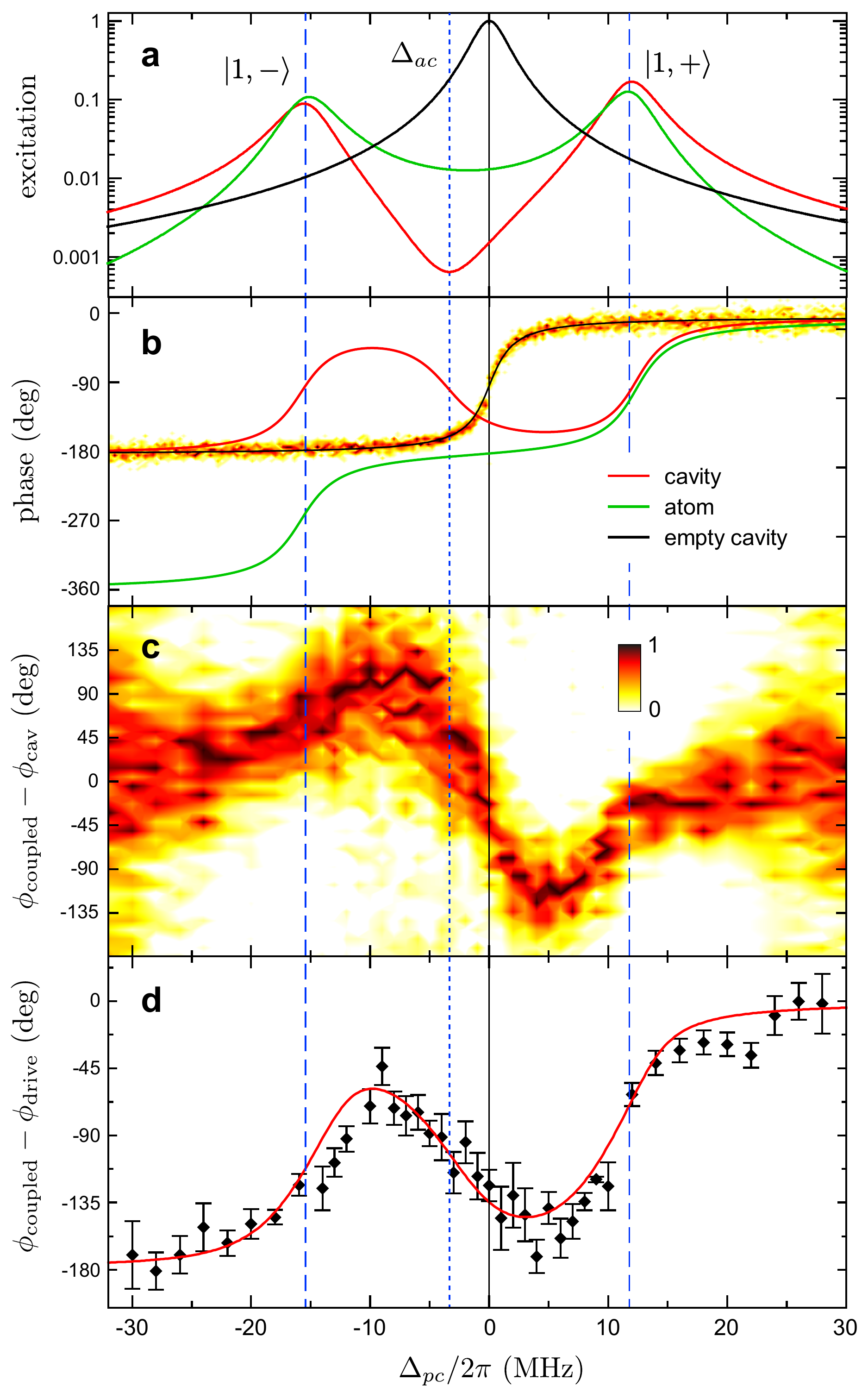}{Phase spectroscopy. (a,b) Theoretical excitation probability (a) and phase (b) of the empty cavity (black) and the two constituents of the strongly coupled system, i.e. the atom (green) and the cavity (red), versus the probe-cavity detuning $\Delta_{pc}$ for our parameters. The experimentally measured phase shift induced by the empty cavity with respect to the driving field is also shown in (b), as a histogram color-coded from white (no events) to black, which is normalized to the maximum number of events for each frequency setting. Vertical dashed lines mark the frequencies of the two normal modes and the dashed line at $-3$\,MHz marks the frequency of the bare atom. (c) Histogram of the additional phase shift caused by an atom strongly coupled to the cavity, referenced to the empty cavity. (d) Measured overall phase shift of the coupled system, derived by adding (b) and (c). The red line is the result of a numerical simulation.}

Phase spectra recorded by heterodyne detection are shown in Figure 2b,c. The phase shift acquired by light transmitted through the empty cavity is overlaid onto the theoretical plot in Fig.\ 2b, and shows the expected arctangent behavior, increasing by $\pi$ as the probe laser is scanned over the resonance. Figure 2c shows the additional phase shift induced by a strongly coupled atom. The sum of the two is the overall phase shift of the coupled system, shown in Fig.\ 2d. Instead of a histogram as in Fig. 2c, the data is here shown as points representing the mean phase shift deduced from fitting to the data a Gaussian distribution that is periodic in phase. The error bars represent the geometric mean of the standard error in the mean and the uncertainty of the mean phase obtained from the fit. The solid red line is the result of a numerical simulation based on Eq.\ (1) which includes effects due to residual atomic motion. The normal-mode resonances can be clearly identified by sharp increases in phase. Between the normal modes, at the antiresonance, an inverse behavior is apparent, with the phase shift exhibiting a negative slope which is maximal at the frequency of the uncoupled atom.

\fig{0.3}{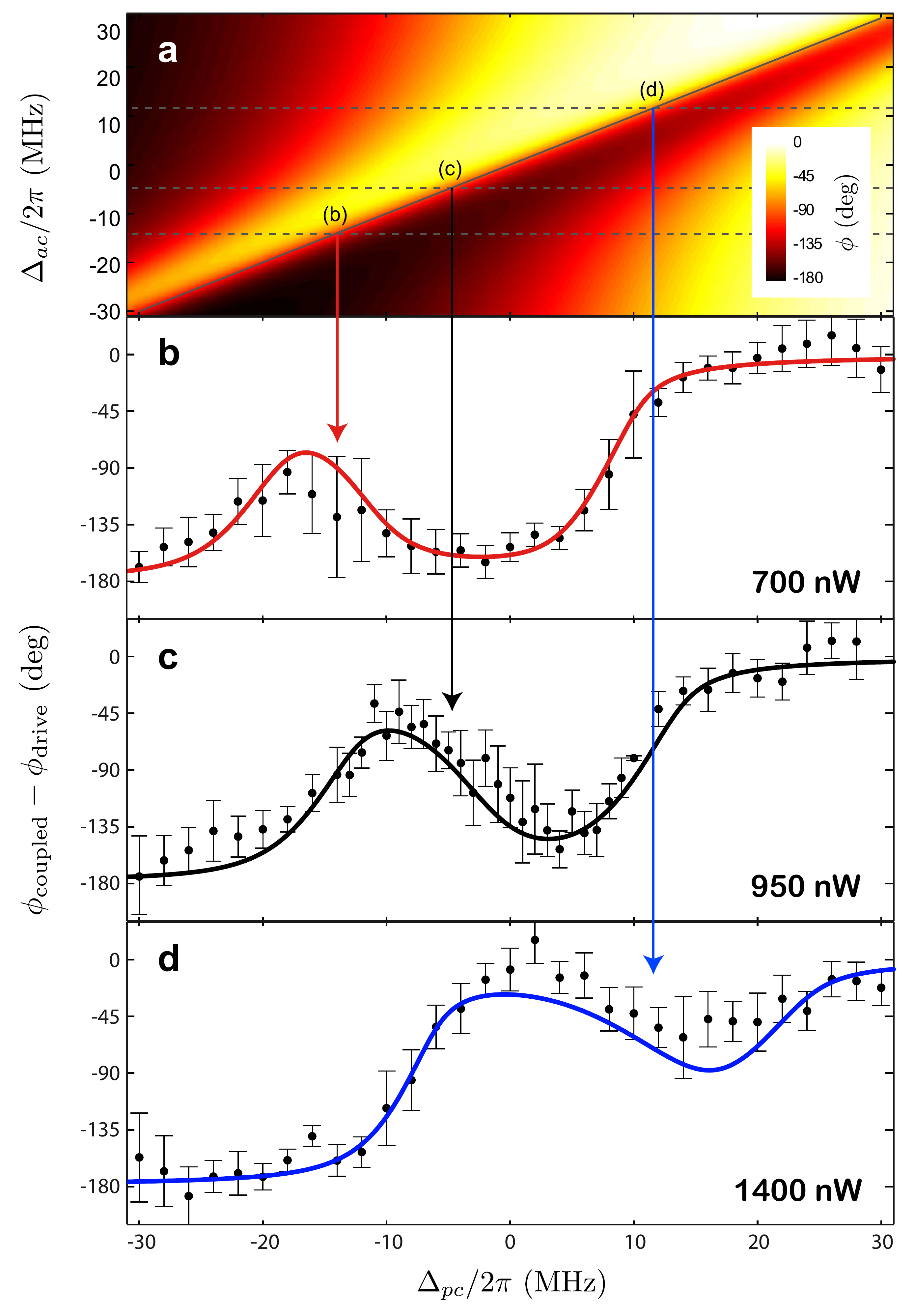}{Tuning of the antiresonance phase shift via the ac-Stark effect. (a) Theoretical phase shift of light transmitted through the strongly coupled system as a function of the probe-cavity $\Delta_{pc}$ (horizontal axis) and atom-cavity $\Delta_{ac}$ (vertical axis) detuning. The diagonal black line indicates where the probe beam is on resonance with the atom. The horizontal dotted lines show the atom-cavity detuning for the scans depicted in the lower plots. Vertical arrows indicate the frequencies of the antiresonances. (b-d) Measured phase shift of the transmitted light for atom-cavity detunings of 12\,MHz (b), $-5$\,MHz (c) and $-14$\,MHz (d), corresponding to dipole-trap laser powers of 1400\,nW, 950\,nW, and 700\,nW, respectively. The solid lines are numerical simulations of the phase shift for each dipole trap laser power.}

The ac-Stark shift induced by the dipole trap light provides a simple way of altering the atom's resonance frequency. In order to verify the behavior depicted in Fig.\ 2, we perform phase measurements across the normal modes for different ac-Stark shifts (i.e.\ different dipole-trap intensities). Figure 3a shows a contour plot of the expected phase as a function of the probe-cavity $\Delta_{pc}$ and atom-cavity $\Delta_{ac}$ detunings. The diagonal line indicates where the probe is resonant with the atom. The horizontal dotted lines mark the atom's detuning at different dipole-trap intensities. The subplots (b-d) show the corresponding measured phase of the light transmitted through the strongly coupled system. The atom is red-detuned from the cavity resonance in (b) and (c), whereas blue detuning is shown in (d). In all scans, the two normal modes are recognizable as positive slopes in the phase on either side of $\Delta_{pc}=0$. The interesting feature, however, is the negative slope of the antiresonance phase shift in between, which always occurs at the atom's resonance frequency (marked with a vertical arrow). This shows that the phase shift indeed directly reflects the frequency of the uncoupled atom.

\fig[b]{0.4}{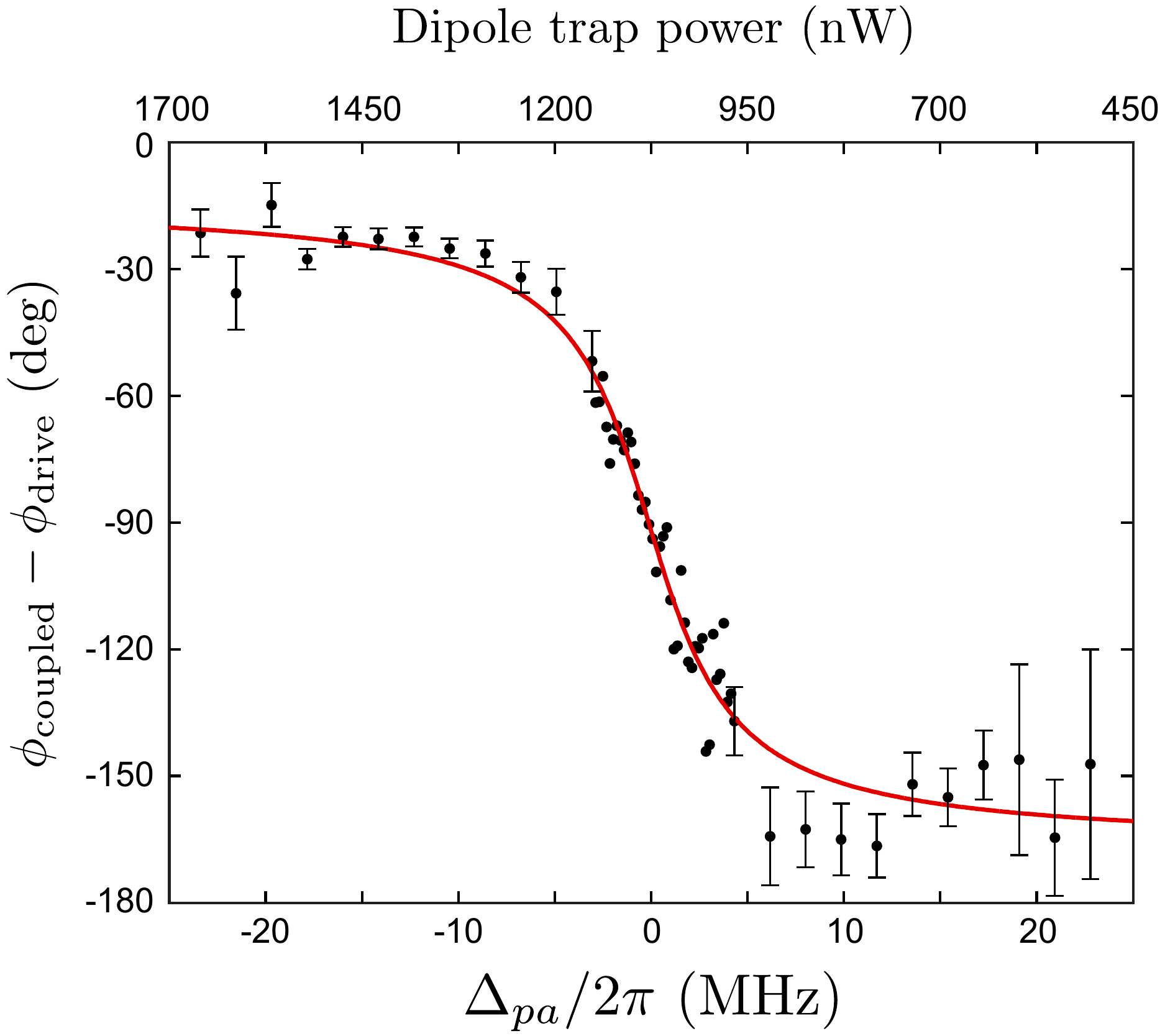}{The phase shift induced by a single atom as a function of probe-atom detuning $\Delta_{pa}$. The probe beam is kept on resonance with the cavity ($\Delta_{pc}$) while the atomic resonance frequency is tuned via the ac-Stark effect induced by the intracavity dipole trap. The dipole trap power is shown on the upper axis. This plot corresponds to a vertical scan in Fig.\ 3 (a). In the central region, error bars are small and omitted for clarity. The larger error bars for $\Delta_{pa}<0$ are caused by the blue detuning of the atom with respect to the cavity, which causes cavity heating \cite{Maunz:2004}. The red line shows an arctangent fit, with a measured width of $3.2\pm0.3$\,MHz that corresponds to the bare-atom decay rate.}

Since the frequency of the antiresonance is exclusively determined by the atom, the ac-Stark shift induced by the dipole trap can be used to optically control the corresponding phase shift. We demonstrate this by measuring the phase shift of the probe light as the dipole power is varied between 450\,nW and 1700\,nW (Fig.\ 4), with the probe laser kept resonant to the empty cavity ($\Delta_{pc} = 0$). As the atom moves across the cavity resonance, we observe a phase shift of 140$^\circ$. This is the largest shift yet observed from a single emitter \cite{Turchette:1995,Fushman:2008,Aljunid:2009,Pototschnig:2011,Jechow:2013}. The theoretical maximum for our system, assuming no atomic motion and maximal coupling to the cavity, is 150$^\circ$. An arctangent fit to the experimental data yields a width of $(3.2 \pm 0.3)$\,MHz, which is in good agreement with the bare atomic decay rate of 3.0\,MHz. This verifies that the atom alone, despite its strong coupling to the cavity, determines the characteristics of the antiresonance phase shift. Moreover, our measurement demonstrates a large ($\sim\pi$) and optically controllable (by means of the dipole trap power) phase shift induced on a single-mode light beam by a single atom.


We now propose to use antiresonance phase shifts for the characterization of complex quantum circuits. Their utility stems from the general result that antiresonances represent what the resonances of the system \emph{would} be if the driven component were held unexcited \cite{Wahl:1999}. This explains why the phase shift in our system has the frequency and width of the atomic resonance, as we drive the cavity mode.

Consider a system of resonators and qubits coupled together in some arbitrary topology (Fig.\ 5a). The excitation spectrum of such a system exhibits distinct resonance-antiresonance behavior under driving (Fig.\ 5b). The resonances depend on properties of all components and their couplings, and are independent of which is driven. The antiresonances, however, depend on everything \emph{except} the component being driven, and therefore provide information about how it affects the total system. By driving each component in turn, information about all of the individual subsystems can be obtained, despite the couplings between them.

As a simple example of this principle, let us suppose that one subsystem exhibits a much larger dissipation than the others, and it is desired to find the lossy component. The system resonances are of no help; their linewidths are an average of the decay rates of all components in the circuit, regardless of which we choose to drive. However, the antiresonances display properties of only the undriven components. Therefore, when the offending component is driven the antiresonances become suddenly narrower, allowing it to be easily identified.

\fig{0.8}{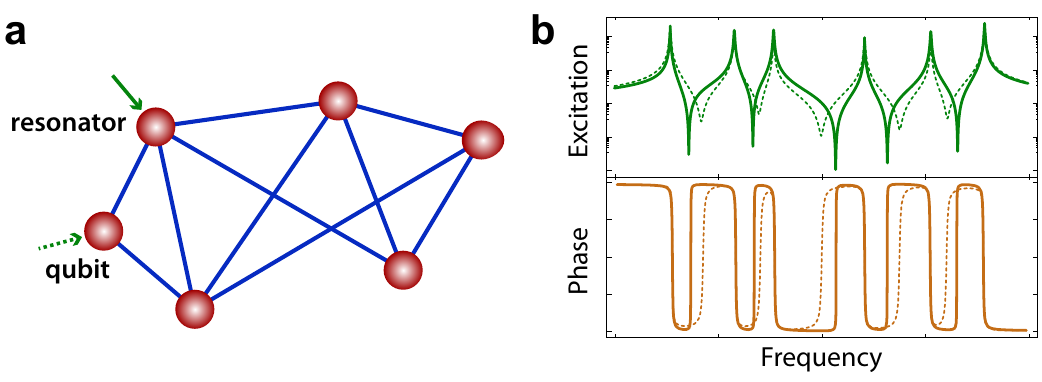}{Antiresonance characterization of complex coupled systems. (a) A notional integrated quantum circuit: the red dots represent circuit components (e.g. qubits or cavities) and the blue lines show the couplings. (b) When driving different components, the system's resonances remain fixed while the positions and widths of the antiresonances change. Measuring the antiresonance phase shifts under different driving conditions therefore facilitates the characterization of the circuit.}


In conclusion, the experimental study carried out here demonstrates a powerful spectroscopic technique that should prove useful in future experiments with interacting quantum systems. In addition, many other potential applications of antiresonances in quantum systems can be envisaged. First, the ability to measure the properties of a single constituent in a strongly coupled system will be valuable in situations where probing the constituents in isolation is impractical, e.g.\ in solid-state cavity QED systems where the emitter and cavity are physically inseparable. Second, the grossly imbalanced distribution of energy among the system constituents at the antiresonance frequency could be useful for cavity cooling of molecules \cite{Horak:1997,Vuletic:2000,Morigi:2007}, since driving the molecules at the empty-cavity resonance frequency would limit their excitation and thus prevent optical pumping into unwanted molecular states. Third, using an emitter with a narrow linewidth may render the antiresonance phase shift useful for optical clock experiments, as it is immune to fluctuations of the cavity. Fourth, nonlinear effects like electromagnetically induced transparency could be incorporated in order to remove the opacity \cite{Mucke:2010,Kampschulte:2010}. The huge phase shift that can be imparted on a light beam by a single emitter might then find an application in quantum-information-processing devices \cite{Turchette:1995}. Finally, our simulations predict giant intensity fluctuations at the cavity-driven antiresonance. One can thus expect large dipole fluctuations for an atom-driven antiresonance. It would be interesting to further explore the connection between these fluctuations and the anomalous atomic momentum diffusion noted by Murr \etal\ \cite{Murr:2006}.


\begin{acknowledgments}
P.~A. would like to thank G.~R.\ Dennis for useful discussions. C.~S.\ acknowledges financial support from the Bavarian Ph.D.\ program of excellence QCCC, and P.~A.\ from the Alexander von Humboldt foundation and the EU through the ITN-project CCQED.
\end{acknowledgments}


\bibliography{antiresonance}

\end{document}